\documentclass[10pt, a4paper]{article} 
\pdfoutput=1
\evensidemargin=\oddsidemargin
\usepackage{amsmath,amssymb,color,graphicx,mathrsfs,cite}
\usepackage{enumitem}
\allowdisplaybreaks
\definecolor{myblue}{RGB}{37,52,148}
\definecolor{myred}{RGB}{207,51,73}
\definecolor{grey}{RGB}{90,90,90}
\definecolor{mygreen}{RGB}{0,86,79}
\usepackage[colorlinks=true,
		    linkcolor=myred,
		    urlcolor=mygreen,
		    citecolor=myblue]{hyperref}
\def\ml{\mathscr L}

\newcommand{\secant}{\text{sec}}

\let\bar=\overline
\def \order(#1){{\mathcal O} \left(#1 \right)}

\def\Eqn#1{Eq.\ (\ref{#1})}
\def\Eqs#1#2{Eqs.\ (\ref{#1}) and (\ref{#2})}

\title	{\LARGE\bf A reappraisal of constraints  on $Z'$ models \\
from unitarity and direct searches at the LHC}

\author {\sf Triparno Bandyopadhyay,$^{a,}
  $\footnote{\href{mailto:gondogolegogol@gmail.com}{\texttt{gondogolegogol@gmail.com}}}~
  Gautam
  Bhattacharyya,$^{b,}$\footnote{\href{mailto:gautam.bhattacharyya@saha.ac.in}
    {\texttt{gautam.bhattacharyya@saha.ac.in}}}~ Dipankar
  Das,$^{a,}$\footnote{\href{mailto:ddphy@caluniv.ac.in}{\texttt{ddphy@caluniv.ac.in}}}~
  Amitava Raychaudhuri$^{a,}$\footnote{\href{mailto:palitprof@gmail.com}{\texttt{palitprof@gmail.com}}}
  \\[10pt] \small\em $^a$Department of Physics, University of
  Calcutta, 92 Acharya Prafulla Chandra Road, Kolkata 700009,
  India\\ \small\em $^b$Saha Institute of Nuclear Physics, HBNI, 1/AF
  Bidhan Nagar, Kolkata 700064, India \\ } \date{\today}

\date{}

\begin{document}
\maketitle	
\begin{abstract}
In a truly model-independent approach, we reexamine a minimal
extension of the Standard Model (SM) through the introduction of an
additional $U(1)$ symmetry leading to a new neutral gauge boson
($Z'$), allowing its kinetic mixing with the hypercharge gauge
boson. An SM neutral scalar is used to spontaneously break this extra
symmetry leading to the mass of the $Z'$.  Except for three
right-handed neutrinos no other fermions are added.  We use the
current LHC Drell-Yan data to put model-independent constraints in the
parameter space of three quantities, namely, $M_{Z'}$, the $Z$-$Z'$
mixing angle ($\alpha_z$) and the extra $U(1)$ effective gauge coupling ($g'_x$),
which absorb all model dependence.  We impose additional constraints
from unitarity and low energy neutrino-electron scattering. However,
limits extracted from direct searches turn out to be most
stringent. We obtain $M_{Z'} > 4.4$ TeV and $|\alpha_z| < 0.001$ at $95\%$ C.L., when
the strength of the additional $U(1)$ gauge coupling is the same as
that of the SM $SU(2)_L$.
\end{abstract}

\section{Introduction \label{sec:intro}} 

Of all the Beyond Standard Model~(BSM) scenarios, none is more
ubiquitous than models with an extra $U(1)$ symmetry in addition
to the SM symmetry, giving a neutral spin-1 massive gauge boson, $Z'$. Its
theoretical motivation comes from various directions.  Left-right
symmetric models, Grand Unified Theories (GUT) larger than
$SU(5)$, e.g. $SO(10)$ or ${E}_6$, as well as string models, all
entail an extra gauged $U(1)$ in addition to the SM group
\cite{lrs1, lrs2, lrs3, lrs4, so101, so102, Gursey:1975ki,
London:1986dk, Langacker:1980js, Rizzo:2006nw, Hewett:1988xc,
Cvetic:1995rj, Leike:1998wr, Langacker:2008yv}.
Non-supersymmetric BSM scenarios, advocated to address the
hierarchy problem, such as 
Little Higgs models
\cite{ArkaniHamed:2001nc,Perelstein:2005ka} with extended gauge
sectors contain $U(1)$ as an extra gauge group. Even dynamical
supersymmetry breaking triggered by an anomalous $U(1)$ has been
extensively discussed (for a review, see \cite{Shadmi:1999jy}).
Leaking of the standard $Z$ boson into an extra dimension yields, from
a four-dimensional perspective, an infinite tower of increasingly more massive
Kaluza-Klein modes, each such mode resembling a $Z'$ boson of a gauged
$U(1)$ carrying specific symmetries
\cite{Antoniadis:1990ew,Appelquist:2000nn,Agashe:2003zs}.  Besides, a
$Z'$ model with a gauged $(B-L)$ symmetry has been used to address the
hierarchy problem by facilitating electroweak symmetry breaking
radiatively {\em \`a la} Coleman-Weinberg keeping classical conformal
invariance and stability up to the Planck scale \cite{Iso:2009nw}.
Cosmological inflation scenarios with non-minimal gravitational
coupling have been studied in a similar context where the inflaton
coupling is correlated to the $Z'$ coupling \cite{Oda:2017zul}.
$U(1)$ gauge bosons also constitute important ingredients in cosmic
string models \cite{Achucarro:1999it}.

On the other hand, $Z'$ has been fruitfully employed in many theoretically
well-motivated models as a portal to dark matter (DM), mediating between the
dark sector and the visible sector \cite{Dudas:2009uq,Dudas:2012pb,Foot:2014uba,
	Foot:2014osa,Okada:2016tci,Okada:2017dqs,Okada:2018ktp}.
The DM itself could be a $U(1)$ gauge boson of the dark sector. A heavy
$Z'$ in such models could be realized in a gauge invariant way by the
St\"uckelberg mechanism \cite{Kors:2004dx}.  In the astrophysical
context too a $Z'$ gauge boson has been advocated to account for the
$\gamma$-ray excess in the galactic center \cite{Berlin:2014tja,Cline:2014dwa}.

Thus there is enough motivation for the $Z'$ mass and coupling to be an
important part of phenomenological studies in the context of colliders
\cite{Carena:2004xs, Rizzo:2006nw, Petriello:2008zr, Salvioni:2009mt,
Accomando:2010fz, Buckley:2011vc, Accomando:2016sge}, the collider-dark
matter interface \cite{Okada:2016gsh, deSimone:2014pda,
Buchmueller:2014yoa, Ducu:2015fda, Klasen:2016qux}, flavor physics
\cite{Gauld:2013qba, Buras:2013qja} and electroweak precision tests
\cite{Erler:2009jh, Bhattacharyya:1991rx, Bhattacharyya:1990bv}.  In this
work we use the latest ATLAS (LHC) Drell-Yan (DY) data (36 ${\rm
fb}^{-1}$ luminosity) to set model-independent bounds on the fermionic
couplings of $Z'$. For this we use the data for both $(e^+e^-$,
$\mu^+\mu^-)$ as well as the $\tau^+\tau^-$ final states. In addition,
we use s-wave unitarity to set upper bounds on $M_{Z'}$ as a function of
the $Z$-$Z'$ mixing angle ($\alpha_z$).  Additionally, we use the low
energy $\nu_\mu$-$e$ scattering data to constrain the $Z'$ parameter
space. The LHC DY data turn out to be most constraining compared to the
other two considerations. {This does not undermine the relevance of
    the other two
    constraints, which have situational merits. The unitarity bound holds
    irrespective of the $Z'$ coupling to fermions, whereas the $\nu_\mu$-$e$
scattering limits become important for hadrophobic $Z'$s.  Taking into
account all the bounds,} we obtain strong constraints in the complete
parameter space spanned by only three independent parameters: $M_{Z'},
\alpha_z$, and $g_x'$, the effective gauge coupling of the additional
$U(1)$ taking into account the scope for kinetic mixing. We make an
important observation that all model dependence can be absorbed within
the above three parameters as long as the additional $U(1)$ is
non-anomalous.

Very recently, constraints directly on $M_{Z'}$ for various $U(1)$
extensions have been derived in \cite{Benavides:2018fzm} using the 36
${\rm fb}^{-1}$ ATLAS data, and wherever we overlap we roughly agree
with their limits.  Constraints directly on $M_{Z'}$ were also
obtained in \cite{Ekstedt:2016wyi} assuming that the $Z$-$Z'$ mixing
angle is small, but those limits are obviously a bit weaker as they
were extracted using the then available ATLAS data with much lower
luminosity.

Our paper is organized as follows.  In Sec.~\ref{sec:general}, we set
up our notations recapitulating the $Z'$-extension of the SM touching
upon the scalar and the fermion sectors. Then, in Sec.~\ref{sec:LHC},
we use the latest 36 ${\rm fb}^{-1}$ ATLAS DY data
\cite{Aaboud:2017buh,Aaboud:2017sjh} to set constraints on its
fermionic couplings for different ${Z'}$ masses in a model-independent
manner.  Next, in Sec.~\ref{sec:uni}, we discuss the bounds on the
$Z'$-mass and the $Z$-$Z'$ mixing angle arising from s-wave
unitarity. Note that this bound depends only on $M_{Z'}$ and the
$Z$-$Z'$ mixing angle and is independent of the $Z'$ couplings to the
fermions.  Once those fermionic couplings are chosen, a bound on the
same plane arises from the low energy $\nu_\mu$-$e$ scattering data,
which we discuss in Sec.~\ref{sec:low}. In Sec.~\ref{sec:results}, we
combine the limits arising from these aspects to identify the region
currently allowed for different $U(1)$ extensions. We end with our
conclusions where we highlight the new features arising out of our
analysis.

\section{Minimal $Z'$ model -- a small recapitulation}
\label{sec:general}
As noted in the introduction, BSM scenarios with an electrically
neutral, massive vector-boson, $Z'$, are quite common in the literature.
The simplest realizations of $Z'$ models are the ones where the SM gauge
symmetry, $\mathcal{G}_{\mathrm{SM}} \equiv SU(3)_C \otimes SU(2)_L
\otimes U(1)_Y$, is minimally extended to $ \mathcal{G}_{\mathrm{SM}}
\otimes U(1)_X$. The $U(1)_X$ is broken by a $\mathcal{G}_{\mathrm{SM}}$
singlet scalar, $S$, charged under $U(1)_X$.  Without any loss
  of generality we choose this charge to be
$1/2$, which fixes the convention for $g_x$ -- the gauge coupling
corresponding to $U(1)_X$.  Thus, in the minimalistic scenario, we have
the following scalar multiplets, transforming under $SU(3)_C\times
SU(2)_L\times U(1)_Y\times U(1)_X$ as:
\begin{eqnarray}
\Phi\equiv (1,2,1/2,x_\Phi/2) \,; \qquad S\equiv 
    (1,1,0,1/2) \,, \label{eq:s_cont}
\end{eqnarray}
where $\Phi$ denotes the usual $SU(2)_L$ doublet responsible for the SM
gauge symmetry breaking as well as the Dirac masses of fermions. The
quantities inside the parentheses characterize the transformation
properties under the gauge group $SU(3)_C\otimes SU(2)_L\otimes
U(1)_Y\otimes U(1)_X$.  The electric charge is given by:
\begin{eqnarray}
    Q&= T_{3L} + Y \,, 
\end{eqnarray}
where $T_{3L}$ and $Y$ are the third component of weak isospin and the
hypercharge respectively. As $\Phi$ transforms in a nontrivial fashion
under $SU(2)_L$, $U(1)_Y$, and $U(1)_X$ there will be mixing among the
neutral gauge boson states when $\Phi$ develops a vacuum expectation
value (vev).  The mass eigenstates which emerge will be identified as
the massless photon $(A)$, the SM $Z$, and an exotic $Z'$.  Note that
even if we start with $x_\Phi=0$, $\Phi$ can develop a $U(1)_X$ charge
due to gauge-kinetic mixing among the two abelian field strength tensors
\cite{Holdom:1985ag}. Also, in general, there will be mixing among the
neutral scalars coming from $\Phi$ and $S$, and a certain
composition of the two should correspond to the SM-like scalar observed
at the LHC. 

%
Abelian extensions of the SM are typically motivated by some high scale
physics related to an elaborate scalar sector, and it might seem that
the two--scalar  scenario we are considering here is a bit too
simplistic. However, we are interested in models where the new physics
beyond the extra $U(1)_X$ is at too high a scale to have any meaningful
contribution to $\mathcal{O}($TeV) physics, or too weakly coupled. With
that in mind, such a minimal framework is capable of describing the
gauge-scalar sector of a wide array of $U(1)$ extensions of the SM,
which are differentiated by the fermionic charges under the $U(1)_X$. In
the following sub-sections, we describe our framework in detail. 
In passing, it should be noted
that in the literature one is often faced with models where the
extended gauge symmetry is given by $SU(3)_C\otimes SU(2)_L \otimes
U(1)_1 \otimes U(1)_2$, where the SM $U(1)_Y$ is a linear combination of
$U(1)_1$ and $U(1)_2$. An example is  $U(1)_R\otimes
U(1)_{\mathrm{B-L}}$, of left-right symmetric models. In such cases, we
can readily perform a rotation among the $U(1)$ generators to obtain
the $U(1)_Y\otimes U(1)_X$ basis that we are using.

\subsection{The gauge-scalar sector}
\label{s:potential}
The gauge-scalar part of the Lagrangian for minimal
$\mathcal{G}_{\mathrm{SM}}\otimes U(1)_X$ models is given by: 
\begin{eqnarray}
    \mathscr{L} = \mathscr{L}_{\mathrm{GK}} + \mathscr{L}_{\mathrm{SK}}
    - V(\Phi,S) \,,
\end{eqnarray}
where $\ml_{\mathrm{GK}}$ and $\ml_{\rm SK}$ are the kinetic Lagrangians
in the gauge and the scalar sectors respectively and $V(\Phi,S)$ denotes the
scalar potential, expressions for which appear below:
\begin{subequations}
    \begin{eqnarray}
        \ml_{\mathrm{GK}} &=&  -\frac{1}{4}W^a_{\mu\nu}W_a^{\mu\nu}
        -\frac{1}{4}B_{\mu\nu}B^{\mu\nu} - \frac{1}{4} X_{\mu\nu}
        X^{\mu\nu} -\frac{\sin\chi}{2} B_{\mu\nu} X^{\mu\nu} \,,
        \label{e:GaugeKd}\\
        \ml_{\rm SK} &=&  (D^\mu\Phi)^\dagger (D_\mu\Phi) +  (D^\mu S)^\dagger
        (D_\mu S) \,, \label{e:ScalK}\\
        V(\Phi,S) &=& -\mu^2(\Phi^\dagger\Phi) - \mu_S^2(S^\dagger S) +
        \lambda_\Phi (\Phi^\dagger\Phi)^2 + \lambda_S (S^\dagger S)^2 +
        \lambda_{\Phi S}(\Phi^\dagger\Phi)(S^\dagger S) \, . 
\label{e:potential}
    \end{eqnarray}
\end{subequations}
Above,
$W^a_{\mu\nu}$, $B_{\mu\nu}$, and $X_{\mu\nu}$ denote the field tensors
corresponding to $SU(2)_L$, $U(1)_Y$, and $U(1)_X$ respectively, and the
covariant derivatives for $\Phi$ and $S$ are given by:
\begin{subequations}
\begin{eqnarray}
D_\mu \Phi &=& \left(\partial_\mu - ig \frac{\tau_a}{2}  W^a_\mu 
-i \frac{g_Y}{2} B_\mu - i \frac{g_x}{2} x_\Phi X_\mu \right)\Phi  \,, \\
D_\mu S &=& \left( \partial_\mu - i \frac{{g}_x}{2}  X_\mu \right)S \, ,
\end{eqnarray}
\end{subequations}
where $\tau_a$ represents the Pauli matrices and the naming  
convention of the gauge fields mirrors that of the field
strength tensors.

Note that, in the ($B^{\mu\nu}, X^{\mu\nu}$) basis, $\ml_{\mathrm{GK}}$
contains the gauge kinetic mixing term $(\sin\chi/2)
B_{\mu\nu}X^{\mu\nu}$ \cite{Holdom:1985ag}. Such a term should, in
general, be present in the lagrangian as it is both Lorentz and gauge
invariant. In a UV complete theory, the parameter $\chi$ should be
calculable by integrating out heavy states at the appropriate scale.
However, we stay blind to such UV completion and treat $\chi$ as a general
parameter. We can perform a general linear transformation to go to a
basis where $\ml_{\mathrm{GK}}$ is canonically diagonal
\cite{Babu:1997st,Brahmachari}:
\begin{eqnarray}
\label{e:nonunitary}
 \begin{pmatrix} {B}_\mu\\{X}_\mu \end{pmatrix}\rightarrow 
\begin{pmatrix} {B'}_\mu\\ {X'}_\mu \end{pmatrix} = 
     \begin{pmatrix} 1 & \sin\chi\\0&\cos\chi\end{pmatrix}
     \begin{pmatrix} {B}_\mu\\{X}_\mu \end{pmatrix} \,.
\end{eqnarray}
In this basis, the gauge-kinetic Lagrangian becomes:
\begin{align}
    \ml_{\mathrm{GK}} &= -\frac{1}{4}W^a_{\mu\nu}W_a^{\mu\nu}
        -\frac{1}{4}B'_{\mu\nu}B'^{\mu\nu} - \frac{1}{4} X'_{\mu\nu}
        X'^{\mu\nu},
        \label{e:GaugeKdiagonal}
\end{align}
and the covariant derivatives take the following forms:
\begin{subequations}
\begin{align}
    D_\mu \Phi &= \partial_\mu\Phi - i \frac{g}{2} \left(\tau_a  W^a_\mu 
+ \tan\theta_w B'_\mu +  \tan\theta_x x'_\Phi X'_\mu \right)\Phi  \,, \\
D_\mu S &= \left( \partial_\mu - i \frac{{g}'_x}{2}  X'_\mu \right)S \,,
\end{align}
\end{subequations}
where we have defined,
\begin{subequations}
\begin{align}
    \tan\theta_w &= \frac{g_Y}{g} \,, \\
    \tan\theta_x &= \frac{g^\prime_x}{g} \,, \label{e:tz} \\
    \mathrm{with}\quad g^\prime_x &= g_x \, \secant\chi  \,,
    \label{e:gzp} \\
    \mathrm{and}\quad x^\prime_\Phi &= x_\Phi - \frac{g_Y}{g_x}\sin\chi 
    \,. \label{e:zp} 
\end{align}
\end{subequations}
\Eqs{e:gzp}{e:zp} reflect how the definitions of the gauge coupling and
the gauge charge of $\Phi$ corresponding to the extra $U(1)$ will be
modified in the presence of kinetic mixing. In the limit of
zero kinetic mixing, $\tan\theta_x$ characterizes the  strength of the 
$U(1)_X$ gauge coupling relative to the weak gauge coupling.

%
After spontaneous symmetry breaking, we expand
the scalar fields, in the unitary gauge, as
\begin{eqnarray}
\Phi = \frac{1}{\sqrt{2}} \begin{pmatrix}
0 \\ v+ \phi_0  \end{pmatrix} \,, \qquad
S =\frac{1}{\sqrt{2}} \left( v_s + s\right) \,,
\end{eqnarray}
where $v$ and $v_s$ are the vevs for $\Phi$ and $S$ respectively.
This will lead to the neutral gauge boson mass matrix, in the basis where the
gauge kinetic terms are diagonal, which can be written as follows:
\begin{eqnarray}
\ml^{\rm mass}_N = \frac{1}{2}
\begin{pmatrix} W^3_\mu  & B^\prime_\mu  & X^\prime_\mu \end{pmatrix}
\cdot {\cal M}_N^2 \cdot
\begin{pmatrix} W^3_\mu  \\ B^\prime_\mu  \\ X^\prime_\mu \end{pmatrix} \,,
\end{eqnarray}
where
\begin{eqnarray}
\label{e:MN2}
{\cal M}_N^2 = \frac{g^2 v^2}{4}
\begin{pmatrix}
1  &  -\tan\theta_w    &  -x^\prime_\Phi \tan\theta_x    \\
-\tan\theta_w  & \tan^2\theta_w & x^\prime_\Phi \tan\theta_x\tan\theta_w \\
- x^\prime_\Phi \tan\theta_x & x^\prime_\Phi \tan\theta_x\tan\theta_w &
\tan^2\theta_x \left(r^2+x^{\prime 2}_\Phi \right)
\end{pmatrix} \,,
\end{eqnarray}
with $r=v_s/v$.
The mass matrix in \Eqn{e:MN2} can be block diagonalized as follows:
\begin{eqnarray}
\label{e:MN2block}
O_w^T \cdot{\cal M}_N^2\cdot O_w = \frac{g^2 v^2}{4}
\begin{pmatrix}
0  &  0    &  0    \\
0  & \secant^2\theta_w & - x^\prime_\Phi \tan\theta_x \secant\theta_w \\
0 & - x^\prime_\Phi \tan\theta_x \secant\theta_w &
\tan^2\theta_x \left(r^2+x^{\prime 2}_\Phi \right)
\end{pmatrix} \,,
\end{eqnarray}
where
\begin{eqnarray}
\label{e:Ow}
O_w = 
\begin{pmatrix}
\sin\theta_w  &  \cos\theta_w    &  0    \\
\cos\theta_w  & -\sin\theta_w & 0 \\
0 & 0 & 1
\end{pmatrix} \,.
\end{eqnarray}
The massless photon, $A_\mu$, is then readily extracted as
\begin{eqnarray}
\label{e:photon}
\begin{pmatrix}
A_\mu \\ Z_{1\mu} \\ X'_{\mu}
\end{pmatrix} = O_w^T
\begin{pmatrix}
W^3_{\mu}  \\ B'_{\mu}  \\ X'_{\mu}
\end{pmatrix} \,.
\end{eqnarray}
Diagonalization of the remaining $2\times 2$ block of
the matrix in \Eqn{e:MN2block} gives rise to the remaining mass
eigenstates, namely, $Z$ and $Z'$. The rotation between the gauge and
the mass bases is given by: %
\begin{eqnarray}
\label{e:gaugerot}
\begin{pmatrix} B^\prime_\mu \\ W^3_\mu \\ X^\prime_\mu \end{pmatrix}
= \begin{pmatrix}
\cos\theta_w & -\sin\theta_w \cos\alpha_z & \sin\theta_w \sin\alpha_z \\
\sin\theta_w & \cos\theta_w \cos\alpha_z & -\cos\theta_w \sin\alpha_z \\
0                  &        \sin\alpha_z              &         \cos\alpha_z
\end{pmatrix}%
\begin{pmatrix} A_\mu  \\  Z_\mu   \\   Z^\prime_\mu \end{pmatrix} \,.
\end{eqnarray}
This second step of diagonalization then entails the following relations:
\begin{subequations}
\label{e:zzp}
\begin{eqnarray}
M_{11}^2 \equiv    M_Z^2 \cos^2{\alpha_z} + M_{Z'}^2 \sin^2\alpha_z
    &=& \frac{M_W^2}{\cos^2\theta_w} \,, \label{e:zzp1}  \\
    M_{Z'}^2 \cos^2{\alpha_z} + M_Z^2 \sin^2\alpha_z
    &=& M_W^2\tan^2\theta_x \left( r^2+x^{\prime 2}_\Phi \right) \,, \label{e:zzp2} \\
    \left( M_{Z'}^2-M_Z^2 \right) \sin 2\alpha_z &=&
    \frac{2 x^\prime_\Phi \tan\theta_x M_W^2}{\cos\theta_w} \,, \label{e:zzp3}
\end{eqnarray}
\end{subequations}
where $M_W = gv/2$ denotes the $W$-boson mass.
  We use \Eqn{e:zzp} to replace $\theta_w$, $r$ and
  $x'_\Phi$ in terms of $M_{Z'}$, $\alpha_z$ and $\tan\theta_x$.
  As we will see later, the latter three quantities  can be extracted
  directly from data in a model-independent way. It is important to
  note that we have not treated $\theta_w$ as the conventional weak
  (Weinberg) angle under the implicit {\em a priori} assumption that
  $\alpha_z$ is small, rather we traded it in favor of $M_{Z'}$ and
  $\alpha_z$ using \Eqn{e:zzp1}.
While the gauge-scalar sector described here holds generally for
minimal $Z'$ models, the fermion charge assignments vary across them.
However, a general formalism can be developed for the fermionic sector
 as well, which we discuss the next
subsection.

\subsection{Anomaly cancellation and fermionic charge
assignments}\label{s:fermion} 
In this work we look at the models in which the fermion sector of the
SM is extended by a right-handed (RH) neutrino, $N_R$, per generation. We
are interested in the situation where the RH neutrinos get Majorana masses
from their Yukawa interactions with $S$. Under the assumption of generation
universality, the possible $U(1)_X$ charge
options for the fermions are quite restricted, as we now discuss.

We assign a $U(1)_X$ charge $x_q$ for the left-handed quark
doublets and $x_l$ for the left-handed lepton doublets. For the
right-handed $u$-type ($d$-type) quarks we assign the charges $x_u$
($x_d$) while for the right-handed electron we take it to be $x_e$. The
$U(1)_X$-charge of the right-handed neutrinos, $N$, is taken as $x_N$. The
$U(1)_X$ quantum numbers of the scalars have already been introduced: the SM
Higgs doublet, $\Phi$, has a charge $x_\Phi/2$, while $S$ has a charge~$1/2$.

Since the scalar $\Phi$ is responsible for the fermion Dirac
masses, we must have
\begin{equation}
x_q-x_u = x_e - x_l = x_d-x_q = -\frac{x_\Phi}{2} \;.
\label{fmass}
\end{equation}
In addition, since $S$ is assumed to be responsible for the Majorana masses of the
right-handed neutrinos, $x_N$ can be determined as
\begin{equation}
x_N = -1/4 \;.
\label{majorana}
\end{equation}
Further, demanding cancellation of gauge and graviational anomalies, we get
\begin{subequations}
\label{e:a1234}
\begin{eqnarray}
\label{a1}
[SU(2)_L]^2 U(1)_X &\Rightarrow& 3 x_q + x_l = 0 \,, \\
\label{a2}
[SU(3)_C]^2 U(1)_X &\Rightarrow& 2 x_q = x_d + x_u \,, \\
\label{a3}
[U(1)_Y]^2 U(1)_X &\Rightarrow& 2x_q + 6 x_l = 16 x_u + 4 x_d +12 x_e \,, \\
\label{a4}
{\rm Gauge ~Gravity}  &\Rightarrow& 6 x_q + 2 x_l = 3 (x_u + x_d) + (x_e + x_N) \,.
\end{eqnarray}
\end{subequations}
It can be checked that the other two constraints that follow from the
$U(1)_Y [U(1)_X]^2$ and  $[U(1)_X]^3$ triangle anomalies are
automatically satisfied.  \Eqn{e:a1234} contains four relations among
the six unknowns $x_q, x_l, x_u, x_d, x_e$, and $x_N$. Taken together
with \Eqn{fmass} and bearing in mind that $x_N$ is fixed from eq.
\Eqn{majorana}, all the $U(1)_X$ charges of the fermions can be determined in
terms of one free parameter\footnote{Ref.~\cite{Ekstedt:2016wyi}
also introduces a parametrization for the $Z'$ fermionic charges, but our
formulation is slightly different.}, $\kappa_x$, as depicted in Table
\ref{t1}.

\begin{table}[!ht]
	\centering
	\begin{tabular}{|c|c|c|c|c|}
		\hline
		Multiplet & $SU(3)_C$ & $SU(2)_L$ & $U(1)_Y$ & $U(1)_X$\\
		\hline
		$Q_L$ & 3 & 2 & 1/6 & $\kappa_x/3$\\
		$u_R$ & 3 & 1 & 2/3 & $4\kappa_x/3-1/4$\\
		$d_R$ & 3 & 1 & -1/3 & $-2\kappa_x/3+1/4$\\
		$L_L$ & 1 & 2 & -1/2 & $-\kappa_x$\\
		$e_R$ & 1 & 1 & -1 & $-2 \kappa_x + 1/4$\\
		$N_R$ & 1 & 1 & 0 & -1/4\\
		$\Phi$ & 1 & 2 & 1/2 & $\kappa_x-1/4$\\
		$S$ & 1 & 1 & 0 & 1/2\\
		\hline
	\end{tabular}
    \caption{\small \em The $U(1)_X$-charge assignments of the
    multiplets, as a function of $\kappa_x$, satisfying the anomaly
    constraints, as well as the transformation properties of the
    multiplets under the SM part of the gauge symmetry.}
	\label{t1}
\end{table}

Different  $U(1)_X$ models are obtained by choosing  $\kappa_x$
appropriately. In Table \ref{t2} we have shown several
alternatives.  For example, the $(B-L)$ extension of the SM
corresponds to $\kappa_x = 1/4$. For this choice the $x$
charges are precisely $(B-L)/4$ -- the overall factor of $1/4$
being a reflection of our chosen normalization of the $U(1)_X$
coupling constant, $g_x$. It is worth noting that for this choice
of $\kappa_x$ the $SU(2)_L$ doublet scalar $\Phi$ has $U(1)_X$
charge $x_\Phi/2 =
0$. Hence, the $Z$-$Z'$ mixing in $B-L$ models is strictly due to
gauge kinetic mixing, which imparts a $U(1)_X$ charge onto
$\Phi$. The choice $\kappa_x = 0$ corresponds to the case where $U(1)_X
\equiv U(1)_R$ under which the left-handed fermions are singlets
while right-handed fermions have charges $\pm 1/4$. The choice
$\kappa_x = 3/20$ gives $U(1)_X \equiv U(1)_\chi$ which
emerges when an $SO(10)$ GUT is broken to $SU(5)
\times U(1)_\chi$.  Finally, with $\kappa_x =
1/5$  we get the $U(1)_R
\times U(1)_{B-L}$ model which can be rotated to the $U(1)_Y
\times U(1)_X$ form with the $U(1)_X$ charge satisfying $5 x =
(B-L)-T_{3R}/2 $.  In Table \ref{t2} we have
also summarized how the usually normalized $U(1)$ charges in these
models are related to the $U(1)_X$ charges given in the last
column of Table \ref{t1}.
\begin{table}[!ht]
    \centering
    \begin{tabular}{|c|c|c|c|c|}
        \hline
        Model & $U(1)_{B-L}$ & $U(1)_{R}$  & $U(1)_\chi$ &
$U(1)_{R} \times U(1)_{B-L}$ \\
        \hline
        &&&&\\
         Charge definitions &$\frac{(B-L)}{4}$&$-\frac{T_{3R}}{2}$ & $-Q_\chi/\sqrt{10}$ &
        $ \frac{1}{5}\left[(B-L) - \frac{1}{2}T_{3R}\right]$\\
        &&&&\\
         \hline
        &&&&\\

$\kappa_x$&$\frac{1}{4}$ & 0 & $\frac{3}{20}$ &$\frac{1}{5}$\\
        &&&&\\
         \hline
    \end{tabular}
	\caption{\small \em $\kappa_x$ for different
     $U(1)_X$ models. Note that for the $B-L$ model, our $U(1)_{B-L}$ charge
     differs from the conventional choice by a factor of $1/4$ due to
     our convention for the gauge coupling of the additional $U(1)_X$.}
	\label{t2}
\end{table}

\subsection{Fermion couplings to gauge bosons}
The parametrization for fermion charges being set, we can now
write down the fermion couplings to $Z$ and $Z'$, which will be
necessary for the subsequent discussions. 
The relevant interaction Lagrangian can be written
as: %
\begin{eqnarray}
\label{e:nu-e-int}
	\ml_{\rm int} = -\frac{g}{2\cos\theta_w} \left[ \bar{f}\gamma^\mu
	\left( g_V^{f}- g_A^{f}\gamma^5 \right) f Z_\mu + \bar{f}\gamma^\mu 
	\left( g_V^{\prime f}- g_A^{\prime f}\gamma^5 \right) f Z'_\mu \right] \,,
\end{eqnarray}
where $f$ stands for a generic fermion. Using the results of Sec. \ref{s:potential} and \ref{s:fermion} we get:

\begin{subequations}
	\label{e:gen-kap}
	\begin{eqnarray}
	g_V^{f} &=& \cos\alpha_z~{\cal G}_V^f + \sin\alpha_z~{\cal H}_V^f \,, ~~~
	g_V^{\prime f} = -\sin\alpha_z~{\cal G}_V^f + \cos\alpha_z~{\cal H}_V^f  \, , \\
	g_A^{f} &=& \cos\alpha_z~{\cal G}_A^f + \sin\alpha_z~{\cal H}_A^f
	\,, ~~~
g_A^{\prime f} = -\sin\alpha_z~{\cal G}_A^f + \cos\alpha_z~{\cal H}_A^f
	\,, 
	\end{eqnarray}
\end{subequations}
where
\begin{equation}
{\cal G}_V^f = -p^f + 2 Q^f \frac{M_W^2}{M_{11}^2} \; , \; 
{\cal H}_V^f = p^f {\cal F} + r^f \frac{M_W}{M_{11}}\tan\theta_x \;, 
\label{e:fermion1}
\end{equation}
and
\begin{equation}
{\cal G}_A^f = T_{3L}^f  \; , \; 
{\cal H}_A^f = -T_{3L}^f {\cal F} + s^f \frac{M_W}{M_{11}}\tan\theta_x \;. 
\label{e:fermion2}
\end{equation}
The quantities $Q^f$ (electric charge), $T_{3L}^f$ (third
component of weak isospin
of $f_L$), $p^f, r^f$, and $s^f$ for the
different fermions are listed in Table \ref{t:coup}.
In Eqs.~(\ref{e:fermion1}) and (\ref{e:fermion2}) ${\cal F}$ is given by
\begin{eqnarray}
{\cal F} \equiv \frac{\left(M_{Z'}^2-M_Z^2 \right)}{M_{11}^2}
\sin\alpha_z\cos\alpha_z \, . ~~~
\label{e:F}
\end{eqnarray}
Through Eqs.~(\ref{e:gen-kap}) to (\ref{e:F}) the fermion
couplings are expressed in terms of measurable quantities and the
chracteristic model-independent constants are given in Table \ref{t:coup}.

For the left-handed neutrinos, for later use, we define
$\kappa_{Z, Z'}$  through  %
\begin{eqnarray}
	g_V^{\nu} = g_A^{\nu} =\frac{\kappa_Z}{2} \,, \qquad
	g_V^{\prime \nu} = g_A^{\prime \nu} =\frac{\kappa_{Z'}}{2} \,.
\label{e:nucoup}
\end{eqnarray}
\begin{table}[!ht]
	\centering
	\begin{tabular}{|c|c|c|c|c|c|}
		\hline
		Fermion $(f)$ & $Q^f$ & $T_{3L}^f$ & $p^f$ & $r^f$ & $s^f$ \\
		\hline
		$u$ & +2/3 & 1/2 & 5/6  & 1/6 & 0\\
		$d$ & -1/3 & -1/2 &-1/6 & 1/6 & 0\\
		$e$ & -1 & -1/2 & -3/2  & -1/2 & 0\\
		$\nu_L$ & 0 & 1/2 &-1/2 & -1/4 & -1/4\\
		$N_R$ & 0 & 0 & 0 & -1/4 & 1/4\\
		\hline
	\end{tabular}
	\caption{\small \em Coefficients entering in the fermionic
          couplings of $Z$ and $Z'$.}
	\label{t:coup}
\end{table}

\noindent 
It is to be noted that the vector and axial-vector couplings of $Z$
and $Z'$ to the fermions depend on three quantities: $M_{Z'}$,
$\alpha_z$ and $\theta_x$.  What is interesting is that $\kappa_x$,
which is a parameter characterizing different models in an
anomaly-free gauged $U(1)_X$ set-up, cancels out for all the
couplings. Curiously, the pre-factor of $\kappa_x$ for each field is
exactly twice its hypercharge (see Table \ref{t1}). The other
contributions to the $U(1)_X$ charges, which depend on $x_N$,
survive. Our choice that the right-handed neutrino, $N_R$, receives
Majorana masses through coupling with $S$ allowed us to set $x_N =
-1/4$.  Since all the observables can be determined in terms of the
three unknowns $M_{Z'}$, $\alpha_z$ and $\theta_x$, our formalism is
completely model-independent, as all model dependence can be soaked
within the above three quantities as long as we stick to an
anomaly-free set-up\footnote{We mention here about the leptophobic
  $Z'$ scenarios (mainly, $E_6$ models) advocated in
  \cite{Babu:1996vt,Chiang:2014yva,Araz:2017wbp}.  Indeed, the
  leptonic couplings of $X'$ can be made to vanish by appropriately
  tuning the kinetic mixing parameter $\chi$. However, the relatively
  heavier mass eigenstate $Z'$ ceases to be truly leptophobic as it
  invariably contains a part of the SM-like weak eigenstate through
  the unavoidably non-vanishing mixing angle $\alpha_z$ in an
  anomaly-free set-up. If instead we force the heavier state $Z'$ to
  be purely leptophobic, we cannot  avoid an untenable corollary
  that $\tan \theta_x = 0$, i.e., the extra $U(1)_X$ gauge coupling
  $g_x$ has to vanish.}.

%

\section{Bounds from direct searches at the LHC}
\label{sec:LHC}
The LHC experiments CMS and ATLAS routinely search for exotic neutral
vector resonances going to $\ell^+\ell^- (\ell\equiv e,\mu, \tau)$
final states (DY modes). The non-discovery of any such new particle
till date translates to exclusion limits on the mass and couplings of
the $Z'$.  In this section we extract such bounds using the latest 36
${\rm fb}^{-1}$ ATLAS data \cite{Aaboud:2017buh}, and cast them in a
model-independent manner.

%
\begin{figure}[!ht]
    \centering
    \includegraphics[width=0.45\textwidth]{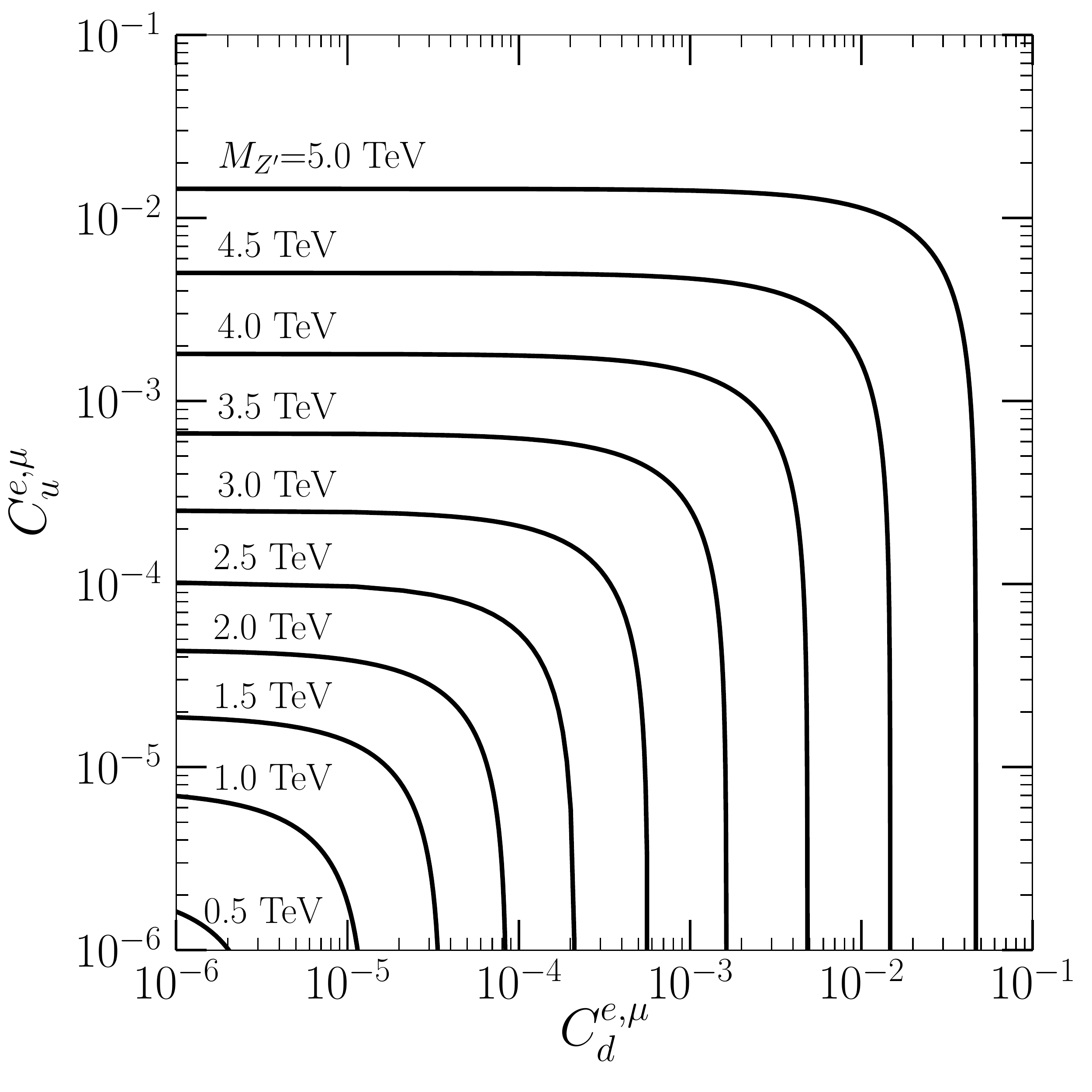}\quad
    \includegraphics[width=0.45\textwidth]{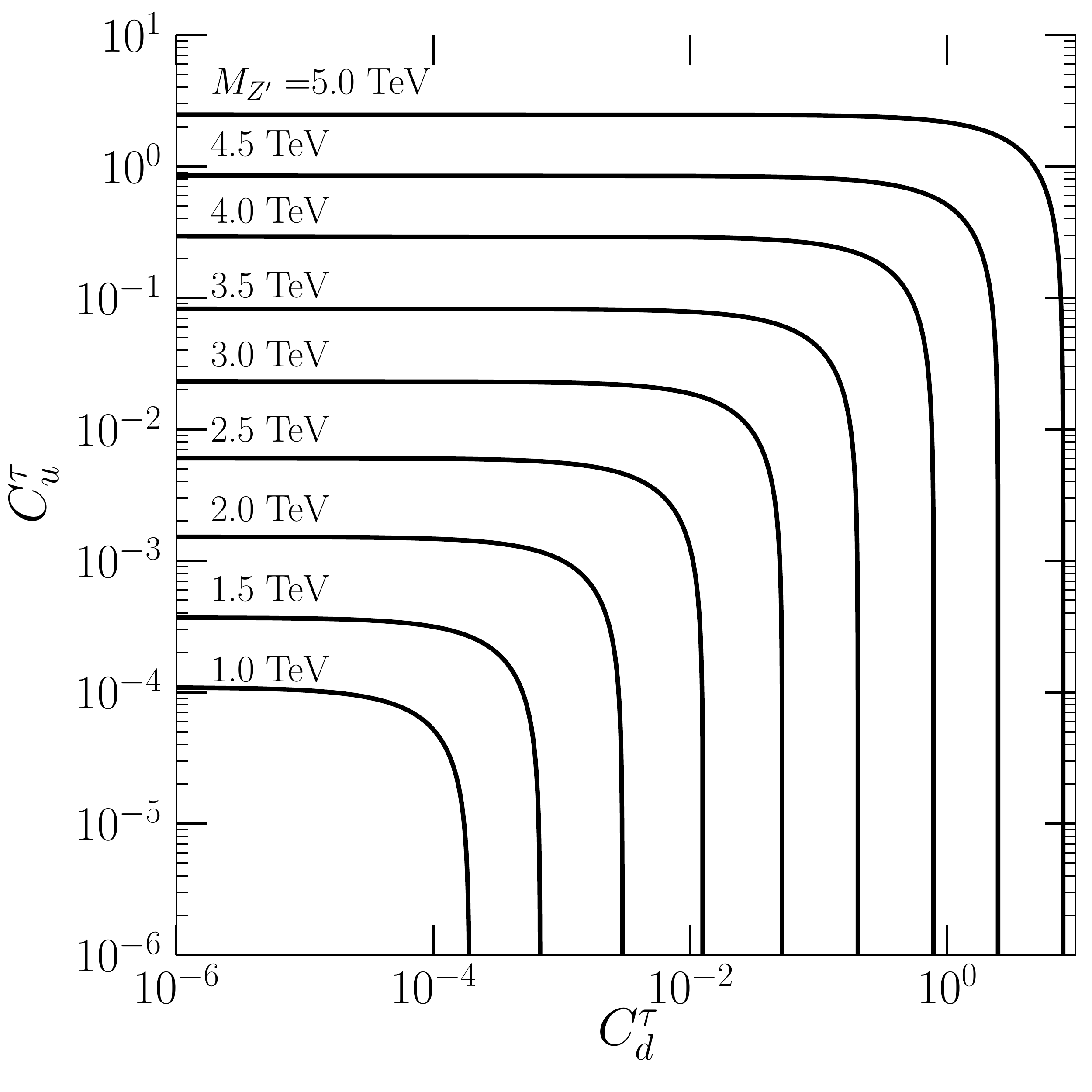}\quad
    \caption{\small \em Exclusion contours at $95\%$ C.L. in the
      $C_u^\ell$-$C_d^\ell$ plane for different values of $M_{Z'}$,
      derived using ATLAS data for dilepton final states
      \cite{Aaboud:2017buh,Aaboud:2017sjh}.  In the left panel the
      contours are for the $\ell \equiv e,\mu$ final state, and the
      right panel corresponds to the $\tau^+\tau^-$ final state.  For
      any given $M_{Z'}$, the interior of the corresponding contour is
      allowed.}
    \label{fig:cucd_ll}
\end{figure}

To analyze the constraints arising from direct resonant $Z'$ production
at the LHC, decaying to a pair of charged leptons, we first define the
chiral couplings $g_L^f$ and $g_R^f$ through:
\begin{equation}
\label{e:LffZp2}
g_R^f = \frac{g}{2\cos\theta_w}(g_V^{\prime f} - g_A^{\prime f}), \;\; 
g_L^f = \frac{g}{2\cos\theta_w}(g_V^{\prime f} + g_A^{\prime f}).
\end{equation}
From \Eqn{e:nucoup} we note that the right-handed couplings of the
light neutrinos to $Z'$, $g_R^\nu$, are zero. In writing
\Eqn{e:LffZp2}, we have implicitly assumed flavor diagonal couplings
for $Z'$, but kept open the possibility of flavor
nonuniversality. With this, the cross section for resonant production
of a $Z'$ boson at the LHC and its subsequent decay into a pair of
charged leptons can be conveniently expressed as (in the narrow width
approximation, for illustration)\cite{Carena:2004xs}:\footnote{The
  reader may notice a difference of a $1/8$ factor between our
  expression and the one given in Ref.~\cite{Carena:2004xs}. This
  issue has been addressed in Refs.~\cite{Feldman:2006wb,Paz:2017tkr}
  whose conventions we follow here.}
\begin{eqnarray}
\label{e:Cq}
	\sigma\left(pp \to Z'X \to \ell^+ \ell^- X\right) =
    \frac{\pi}{6s}\sum_{q} C_q^\ell w_q\left(s,M_{Z'}^2\right) \,,
    \label{eq:CS}
\end{eqnarray}
where the sum is over all the partons. 
The co-efficients,
\begin{eqnarray}
\label{e:Cql}
	C_q^\ell = \left[\left(g_L^q\right)^2 +
	\left(g_R^q\right)^2\right] {\rm BR}\left(Z' \to
	\ell^+ \ell^- \right)
\end{eqnarray}
involve the fermionic couplings of $Z'$ and hence depend on the
details of the fermionic sector of the model under consideration. The
functions $w_q$, on the other hand, contain all the information about
the parton distribution functions~(PDFs) and QCD corrections, detailed
expressions for which appear in the Appendix. Considering the fact
that $w_u$ and $w_d$ are substantially larger than the $w_q$ functions
for the other quarks, we can approximate \Eqn{e:Cq} as
follows\footnote{For most $Z'$ models this is a reasonable
  approximation. In particular, in models with flavor universal $Z'$
  couplings we have checked that it hardly makes a visible difference
  if we use \Eqn{e:Cq} instead of the approximate formula of
  \Eqn{e:cucd_one}.  But, of course, this approximation breaks down in
  the extreme case when the $Z'$ does not couple at all to the first
  generation of quarks\cite{Andrianov:1998hx}.}:
\begin{eqnarray}
\label{e:cucd_one}
	\sigma\left(pp \to Z'X \to \ell^+ \ell^- X\right) \approx
	\frac{\pi}{6s} \left[C_u^\ell w_u\left(s,M_{Z'}^2\right)
	+C_d^\ell w_d\left(s,M_{Z'}^2\right) \right] \,.
\end{eqnarray}

%

Direct searches at the LHC put upper limits on the left-hand-side of
\Eqn{eq:CS}. The most recent ATLAS limits can be found in
\cite{Aaboud:2017buh,Aaboud:2017sjh} where, as expected, the bound for
the $\ell^\pm \equiv \tau^\pm$ case is less stringent than for $\ell^\pm
\equiv e^\pm, \mu^\pm$. Using the CT14NLO PDF set \cite{Dulat:2015mca},
we evaluate $w_u$ and $w_d$, and translate the limit on the cross
section into a bound in the $C_u^\ell$-$C_d^\ell$ plane for different
values of $M_{Z'}$. The results have been displayed in
Fig.~\ref{fig:cucd_ll}, where the left panel corresponds to $\ell\equiv
e,\mu$,\footnote{Such an analysis was carried out by CMS using their
8~TeV (20~fb$^{-1}$) dilepton data\cite{Khachatryan:2014fba}. A
comparison with our results shows that there is almost an order of
magnitude improvement in the corresponding bounds, if we use the current
13~TeV (36~fb$^{-1}$) data.  } and the right panel corresponds to
$\ell\equiv \tau$. For any chosen $M_{Z'}$, only the interior of the
corresponding contour is allowed. Although the bound arising from the
$\tau^+ \tau^-$ final state is substantially weaker compared to that
from $e^+ e^-, \mu^+\mu^-$ final state, it may have its own advantage
for scenarios where, {\it e.g.}, the $Z'$ dominantly couples to the
third generation of fermions\cite{Muller:1996dj, Lynch:2000md,
Benavides:2016utf}.

%
%

\section{Theoretical constraint from unitarity}
\label{sec:uni}
{For $U(1)$ extended models, in the absence of a
$Z'$,} the scattering amplitude for the process $W_L^+W_L^- \to
W_L^+W_L^-$, where $W_L^\pm$ denotes the longitudinal component of the
$W$-boson, will grow as the fourth power of the center of momentum~(CoM)
energy at the leading order. To put it explicitly, if the $Z'$ is too
heavy to contribute, then we can write the Feynman amplitude for
$W_L^+W_L^- \to W_L^+W_L^-$ as
\begin{eqnarray}
\label{e:WWWW}
{\mathscr M}_{W_L^+W_L^- \to W_L^+W_L^-}
= \frac{g^2\cos^2\theta_w E^4}{M_W^4} \sin^2\alpha_z 
\left( -3+6\cos\theta +\cos^2\theta \right) +\order(\frac{E^2}{M_W^2}) \,,
\end{eqnarray}
where $E$ denotes the CoM energy and $\theta$ is the scattering angle. 
From \Eqn{e:WWWW}, the $l =0$ partial wave amplitude which usually
gives the strongest bound, can be extracted as
\begin{eqnarray}
a_0 = -\frac{8}{3} \frac{g^2\cos^2\theta_w E^4}{M_W^4} \sin^2\alpha_z \,.
\end{eqnarray}
Unitarity restricts the magnitude of $a_0$ as $|a_0|< 8\pi$ which translates
into an upper bound for the CoM energy,
\begin{eqnarray}
E < E_{\rm max} = \left[8\pi \times \frac{3\left(
M_Z^2 \cos^2{\alpha_z} + M_{Z'}^2 \sin^2\alpha_z \right)}{
32\sqrt{2} G_F \sin^2\alpha_z } \right]^\frac{1}{4}  \,,
\end{eqnarray}
where $G_F$ is the Fermi constant obtained via the relation,
\begin{eqnarray}
\label{e:GFMW}
g^2/M_W^2 = 4\sqrt{2} G_F \,,
\end{eqnarray}
and we have used \Eqn{e:zzp1} to substitute for $M_W^2/\cos^2\theta_w$. Thus,
to restore unitarity, effects of the $Z'$ must set in before the CoM energy
reaches $E_{\rm max}$, {i.e.}, $M_{Z'}< E_{\rm max}$ which implies:
\begin{eqnarray}
\label{e:unibound}
	\frac{M^4_{Z'} \sin^2\alpha_z}{\left(M_Z^2 \cos^2{\alpha_z} +
          M_{Z'}^2 \sin^2\alpha_z \right)} < 8\pi \times
        \frac{3}{32\sqrt{2}G_F} \,.
\end{eqnarray}

To find a physical interpretation for the above bound, we write down the expression
for the $Z'\to W^+W^-$ decay width as
\begin{eqnarray}
\label{e:ZpWW}
\Gamma(Z'\to W^+W^-) \approx \frac{1}{64\pi} 
\frac{g^2\cos^2\theta_w \sin^2\alpha_z}{3} M_{Z'}
\left(\frac{M_{Z'}}{M_W} \right)^4 \,,
\end{eqnarray}
which is valid in the limit $M_{Z'}\gg M_W$ when the longitudinal
components of the $W$-bosons
dominate\cite{delAguila:1986ad,Deshpande:1988py}.  Substituting for
$\cos\theta_w$ using \Eqn{e:zzp1}, one can easily verify that this
partial decay width increases with $\sin\alpha_z$ as well as $M_{Z'}$.
However, the resonance should be narrow enough so that it can be
distinguished experimentally from the flat background.  In view of
this, it may be reasonable to impose a rather conservative limit,
\begin{eqnarray}
\label{e:decaylimit}
\Gamma(Z'\to W^+W^-) < M_{Z'} \,.
\end{eqnarray}
Using \Eqs{e:zzp1}{e:GFMW} one can check that the above bound can
be translated into
\begin{eqnarray}
\label{e:decbound}
	\frac{M^4_{Z'} \sin^2\alpha_z}{\left(M_Z^2 \cos^2{\alpha_z} +
          M_{Z'}^2 \sin^2\alpha_z \right)} < 48\pi \times
        \frac{1}{\sqrt{2}G_F} \,,
\end{eqnarray}
which is slightly weaker than the unitarity bound in \Eqn{e:unibound}. Therefore,
consideration of unitarity implicitly keeps the corresponding partial decay
width under control.\footnote{
It is worth remarking that such a lesser known virtue of the unitarity bound is
also present in the case of the SM Higgs boson. For $m_h \gg M_W$,
$\Gamma(h^{\rm SM}\to W^+W^-)$ grows as $m_h^3$ and would equal
$m_h$ for $m_h\approx 1.4$~TeV\cite{Gunion:1989we}. But the bound
$m_h<1$~TeV from the $W_L^+W_L^-$ scattering ensures that such a
situation never arises.
}

\begin{figure}[htpb]
    \begin{minipage}[t]{0.46\textwidth}
        \centering
        \vspace{0pt}
        \includegraphics[width=\textwidth]{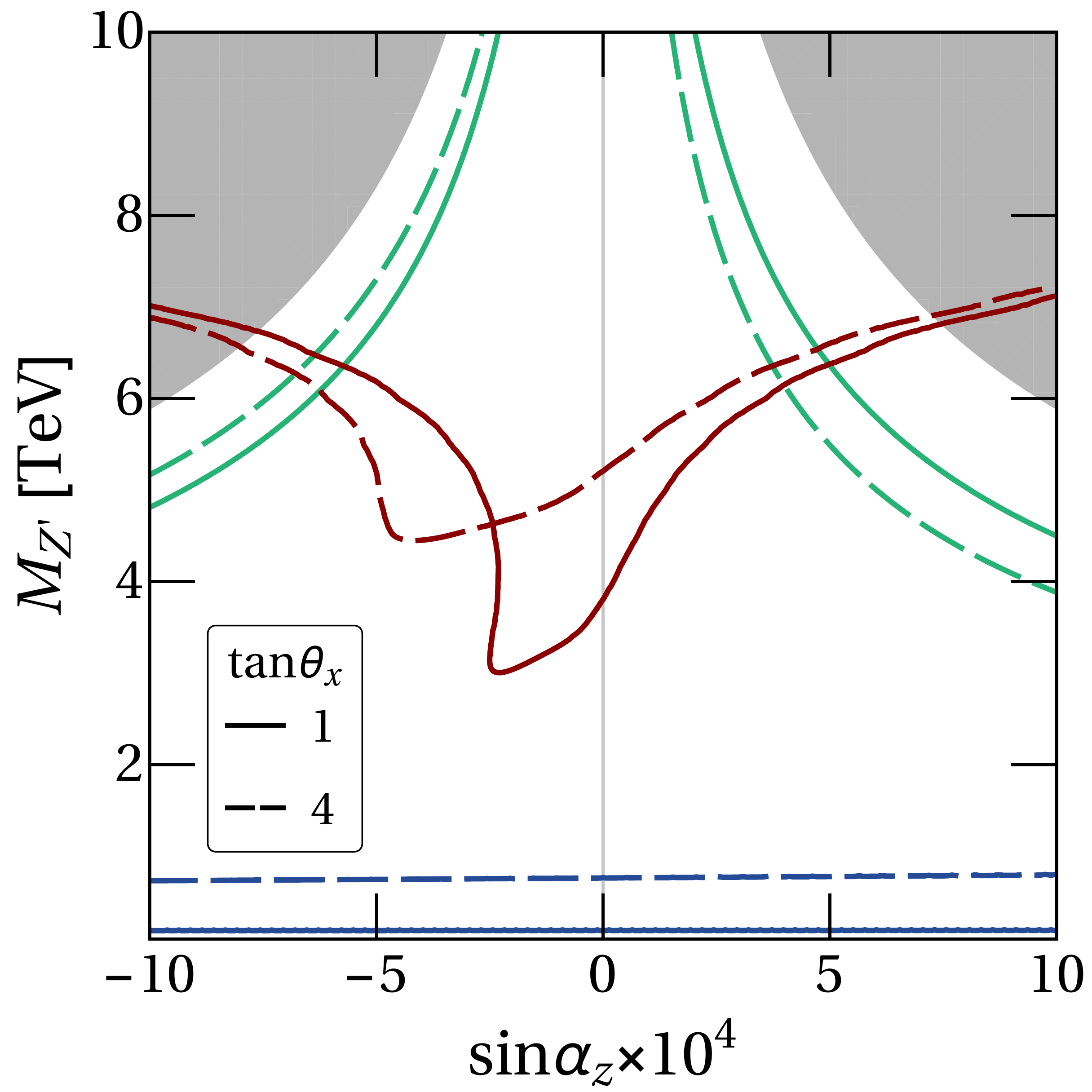}
        \caption{\small \em Consolidated bounds in the
          $(\sin\alpha_z$-$M_{Z'})$ plane for anomaly-free $U(1)_X$
          models. The shaded region is excluded from unitarity. The red and
          the blue colors indicate the limits set by direct detection and
          $\nu_\mu$-$e$ scattering data, respectively. The green contours are
          obtained by setting $\Gamma_{Z'}= M_{Z'}/2$. The solid and dashed
          line-types correspond to $\tan\theta_x= 1$ and $4$,
          respectively. Region above the red lines are allowed by the 36 ${\rm
            fb}^{-1}$ ATLAS data, whereas the region above the blue lines and
          the interior of the green contours represent the allowed area from
          the $\nu_\mu$-$e$ scattering data and $\Gamma_{Z'}\le M_{Z'}/2$,
          respectively.}
        \label{fig:contours}
    \end{minipage}\qquad
    \begin{minipage}[t]{0.46\textwidth}
        \centering
        \vspace{0pt}
        \includegraphics[width=\textwidth]{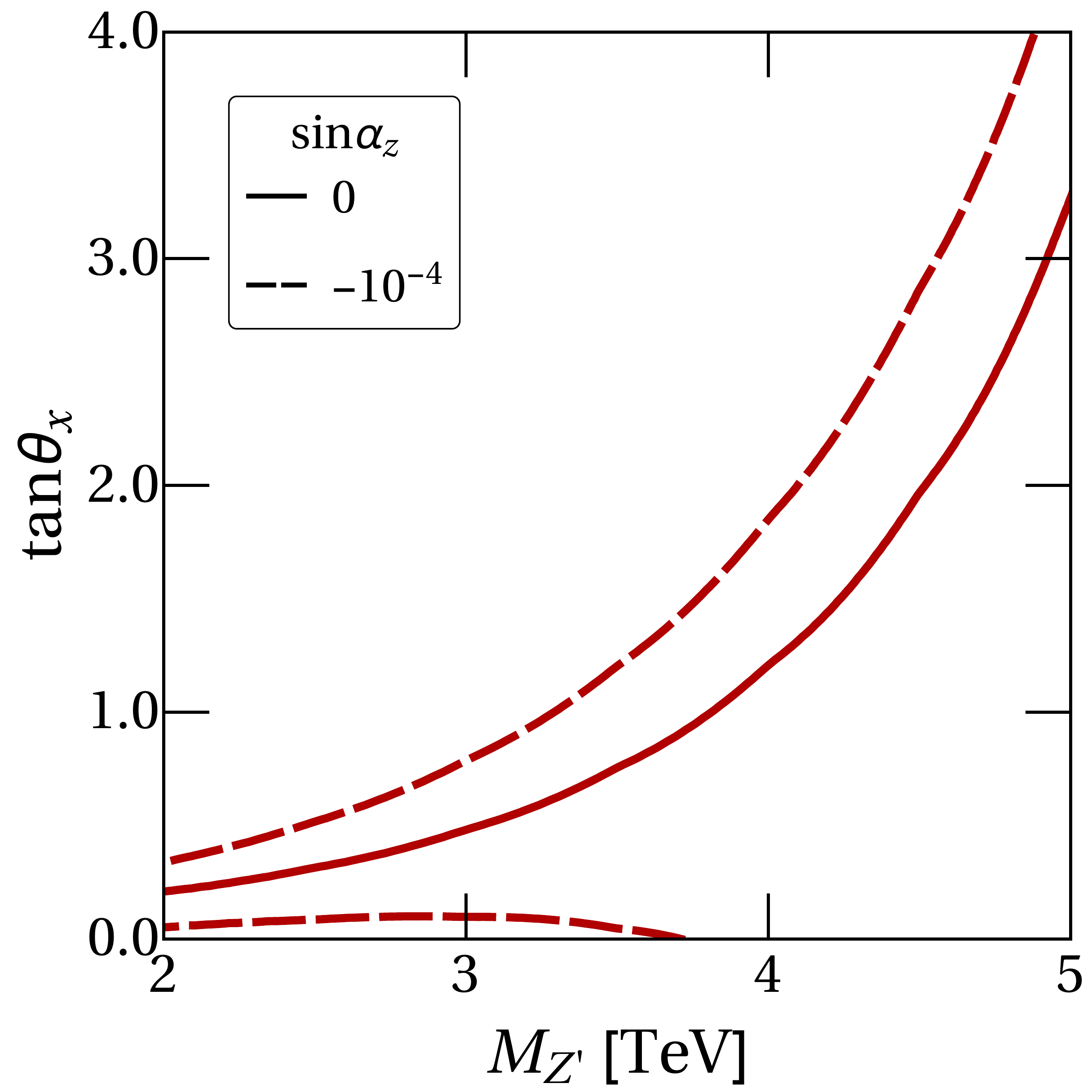}
        \caption{\small \em Bounds in the
          $(M_{Z'}$-$\tan\theta_x)$ plane for anomaly-free $U(1)_X$ models
          using two representative values of $\sin\alpha_z$, namely, $0$
          and $(-10^{-4})$. For these choices of $\sin\alpha_z$ the
          strongest limits arise from the direct searches, which have been
          displayed as the red lines. For $\sin\alpha_z=0$ the region to
          the right of the solid red curve is allowed, whereas for
          $\sin\alpha_z=(-10^{-4})$ the allowed region lies within the
          dashed red curves.}
        \label{fig:gymz}
    \end{minipage}
\end{figure} 

The tree unitarity constraint is of prime importance as it translates
to an {\emph{upper bound}}  on $M_{Z'}$, for a given
$\sin\alpha_z$, {complementing the lower bound that comes from direct
search experiments.} This can be
seen from \Eqn{e:unibound}\footnote{Similarly for $f\bar{f}\to
  W_L^+W_L^-$ the scattering amplitude will grow as $\order(E^2)$
  \cite{Bhattacharyya:2012tj} and can give an upper bound on $M_{Z'}$
  for nonzero $\alpha_z$. But this bound will depend on the fermionic
  couplings of $Z'$\cite{Babu:2011sd} and will not be as model
  independent.}. We show this explicitly when we discuss the interplay
of the different bounds in Sec. \ref{sec:results}. It should also be
noted that although unitarity in the context of $Z'$ models have been
studied earlier \cite{Cheung:2009um,Radel:1993sw}, to our knowledge, the
possibility of using it to cast an upper bound on the $Z'$ mass as in
\Eqn{e:unibound} has not been emphasized before and thus constitutes a new
observation in our paper.  Moreover, since this analysis does not
depend on the details of the fermionic couplings, such a bound is quite
general and can be applied to a wide class of $Z'$ models.

\section{Constraints from  $\nu_\mu$-$e$  scattering}
\label{sec:low}
The unitarity constraint, described in the previous section, relies on
sniffing the effects of $Z'$ through the $Z$-$Z'$ mixing. Therefore,
the bounds are lifted in the limit $\sin\alpha_z=0$ as has been
clearly depicted in Fig.~\ref{fig:contours}.  However, depending on
how $Z'$ couples to the fermions, it is possible to put lower bounds
on $M_{Z'}$, even in the limit of vanishing $Z$-$Z'$ mixing
\cite{Lindner:2018kjo,Abdullah:2018ykz}. This can be done, {\it e.g.},
by using the data from low energy neutrino-electron scattering such as
$\nu_\mu e\to \nu_\mu e$ which proceeds at the tree level purely via
neutral current {(see, \emph{e.g.},
\cite{Hasert:1973cr,Radel:1993sw,Williams:2011qb})}. In models with an extra $U(1)$, the $Z'$ boson will,
in general, also contribute to the scattering.

The dimension-six operator governing $\nu_\mu$-$e$ scattering at low
energies is written as:
\begin{eqnarray}
\label{e:g-nu-e1}
	\ml_{\nu e} = -\frac{G_F}{\sqrt{2}} \left[\bar{\nu}\gamma^\mu 
	\left(1-\gamma^5\right) \nu \right] \left[ \bar{e}\gamma_\mu 
	\left( g_V^{\nu e}- g_A^{\nu e}\gamma^5 \right) e \right] \,.
\end{eqnarray}
We recall that in the SM, the expressions for $g_V^{\nu e}$ and $g_A^{\nu e}$
are very simple at the tree level and are given by
\begin{eqnarray}
	\left(g_V^{\nu e} \right)^{\rm SM} \equiv \left(g_V^{e} \right)^{\rm SM}
	= -\frac{1}{2}+ 2\sin^2\theta_w \,, \qquad
	\left(g_A^{\nu e} \right)^{\rm SM} \equiv \left(g_A^{e} \right)^{\rm SM}
	= -\frac{1}{2} \,.
\end{eqnarray}
Of course, in the $Z'$ models under consideration, the above
expressions will be modified (see \Eqn{e:nucoup}) as follows:
\begin{eqnarray}
\label{e:g-nu-e-model1}
		\left(g_{(V,A)}^{\nu e} \right)^{\rm model} &=& M_{11}^2 \left(
		\frac{\kappa_Z g_{(V,A)}^{e}}{M_Z^2} + \frac{\kappa_{Z'} 
			g_{(V,A)}^{\prime e}}{M_{Z'}^2} \right) \,,
\end{eqnarray}
where the expression for $M_{11}^2$ appears in \Eqn{e:zzp1} and
the rest of the couplings in \Eqn{e:gen-kap}.

We use this formula along with the following {global fit values from
PDG \cite{Patrignani:2016xqp}:}
\begin{eqnarray}
g_V^{\nu e} = -0.040 \pm 0.015 \,, \qquad
g_A^{\nu e} = -0.507 \pm 0.014 \,,
\end{eqnarray}
to draw the $2\sigma$ allowed regions in the $\sin\alpha_z$-$M_{Z'}$
plane for two different values of $\tan\theta_x$ as shown by the blue
curves in Fig.~\ref{fig:contours}.

\section{Results and discussions}
\label{sec:results}
Till now we have developed a general formalism on how to constrain a
minimal $Z'$ model from theoretical considerations as well as from
different types of experimental data.  Now we combine the different
limits together, described in the previous sections, to obtain stronger
bounds on the parameter space. 
To illustrate, $C_{u,d}^{e,\mu}$ and $g_V^{\nu e}$\footnote{ Using the
  expressions in \Eqn{e:gen-kap}, we have checked that $g_A^{\nu
    e}=-0.5$ is independent of the model parameters.}  can be
determined, using Eqs. (\ref{e:Cql}), (\ref{e:LffZp2}) and
(\ref{e:g-nu-e-model1}) in conjunction with \Eqn{e:gen-kap}, in terms
of the three quantities $M_{Z'}$, $\alpha_z$ and $\tan\theta_x$.  The
bound from the left panel of Fig.~\ref{fig:cucd_ll} and the constraint
coming from $\nu_\mu$-$e$ scattering can then be translated to the
limits on those three parameters.

In Fig.~\ref{fig:contours} these bounds have been displayed in the
$\sin\alpha_z$-$M_{Z'}$ plane for any anomaly-free $U(1)_X$ model for
two typical choices of $\tan\theta_x$.  The region excluded from
unitarity has been shaded in gray and is independent of
$\tan\theta_x$. The lower bounds on $M_{Z'}$, arising from the ATLAS
(13 TeV, 36 ${\rm fb}^{-1}$) exclusion of the DY production of $Z'$,
are depicted as red curves, whereas the region above the light blue
curves denote the region consistent with $\nu_\mu$-$e$ scattering.
{Additionally,
    we also give contours that represent a constraint on the $Z'$ decay
    width, as a guideline for the validity of a particle
    interpretation. The green lines in the figure arise from
    the consideration\footnote{What constitutes an
    acceptable width of a heavy particle, or how far the narrow
    width approximation holds good can be a matter of discussion and hence
    we choose to veer on the conservative side, to illustrate what role
    the consideration of width might play in restricting the
parameter space.} $\Gamma_{Z'}\le M_{Z'}/2$.} 

For all the colored contours, the solid
(dashed) curves correspond to $\tan\theta_x=1 (4)$. Recall that
$\tan\theta_x$ is proportional to the effective $U(1)_X$ coupling,
$g'_x$. 
%
As it happens, the lower bounds on $M_{Z'}$ arising from low-energy
$\nu_\mu$-$e$ scattering are considerably weaker than those from direct
searches. However, $\nu_\mu$-$e$ scattering can put important
constraints for hadrophobic $Z'$ models when the production of the $Z'$
at the LHC is very suppressed.  Combining the lower bound on $M_{Z'}$
from the direct searches with the corresponding upper bound coming from,
{\it e.g.}, unitarity, we are able to extract an upper limit on the
magnitude of the $Z$-$Z'$ mixing angle, $\alpha_z$.  Such bounds on
$|\alpha_z|$ are at par with the corresponding limits from electroweak
precision data \cite{Czakon:1999ha,Erler:2009jh}.

%
\begin{table}[htbp]
    \centering
    \begin{tabular}{|l|l| c |}
        \hline 
        &Maximum $|\sin\alpha_z|$ & $10^{-3}$   \\
        $\tan\theta_x=4$ &$M_{Z'}$ exclusion at $\alpha_z=0$ [in TeV] 
        & 5.1   \\
        &Lowest possible value of $M_{Z'}$ [in TeV]
        & 4.4 \\
        \hline
        \hline
        &Maximum $|\sin\alpha_z|$ & $10^{-3}$   \\
        $\tan\theta_x=1$ & $M_{Z'}$ exclusion at $\alpha_z=0$ [in TeV] 
        & 3.8 \\
        &  Lowest possible value of $M_{Z'}$  [in TeV]
        & 3.0 \\
        \hline
    \end{tabular}
    \caption{\small \em Summary of bounds on $M_{Z'}$ and $\alpha_z$
      for anomaly-free $U(1)_X$ models using two representative values
      of $\tan\theta_x$ (which is proportional to the effective
      $U(1)_X$ coupling).}
    \label{t3}
\end{table}
%

%
In Table \ref{t3}, we have summarized the bounds on $\alpha_z$ and
$M_{Z'}$ for $\tan\theta_x$=1 and 4 for anomaly-free $U(1)_X$
models. From Table~\ref{t2} we recall that the choice $\tan\theta_x=4$
corresponds to $g'_x=g$ for the `conventional' $(B-L)$ model. This is
so because for $(B-L)$ model in our normalization, $\kappa_x = 1/4$,
and $g'_x \kappa_x$ in our setup is equivalent to a generic $g'_x$ in
the conventional $(B-L)$ model.  It should be pointed out that
although we have taken into account the decays $Z'\to W^+W^-$ and
$Z'\to Zh$ ($h$ being the lighter SM-like Higgs scalar) for our
analysis, we have assumed the decays $Z'\to NN$, where $N$ denotes a
heavy RH neutrino, and $Z'\to ZH$, where $H$ is the heavier
nonstandard scalar, to be kinematically forbidden. The lower bound on
$M_{Z'}$ is likely to be diluted further if these decay channels open up.

\begin{figure}[htpb]
    \begin{minipage}[t]{0.47\textwidth}
    \vspace{0pt}
\centerline{\includegraphics[width=\textwidth]{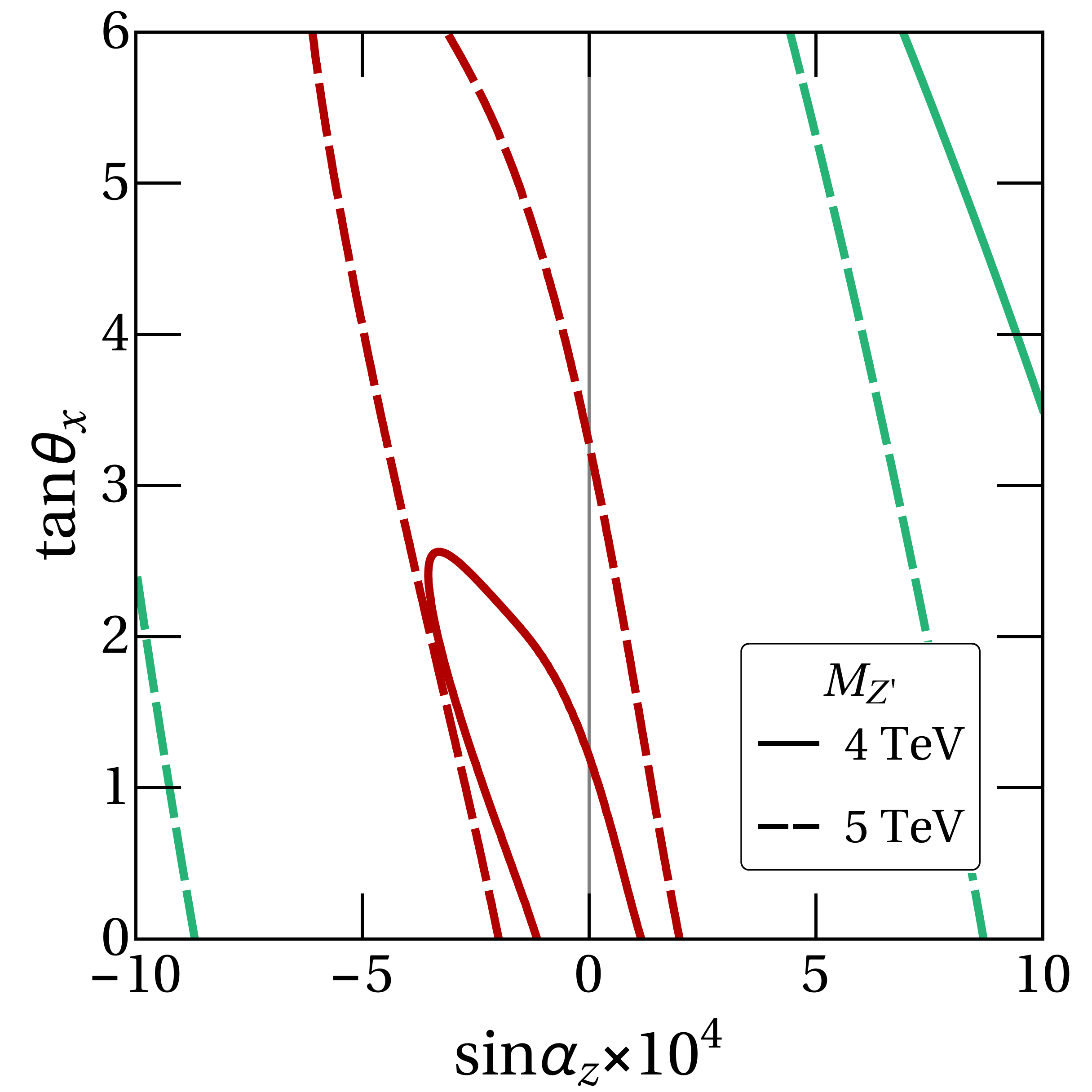}}
\caption{\small \em Consolidated bounds in the
    $(\sin\alpha_z$-$\tan\theta_x)$ plane for anomaly-free $U(1)_X$
    models using two representative values of $M_{Z'}$, namely, $4$~TeV
    and $5$~TeV. For these values of $M_{Z'}$ strongest limits come from
    direct searches, displayed by the red lines. For $M_{Z'}=4$~TeV the
    region inside the solid red contour is allowed, whereas for
    $M_{Z'}=5$~TeV the region bounded within the dashed red lines is
    allowed. 
The green lines refer to $\Gamma_{Z'}\le M_{Z'}/2$ for which the limits
are rather weak for the chosen values of $M_{Z'}$ (region inside the
dashed lines are allowed for $M_{Z'} = 5$ TeV, while for $M_{Z'} = 4$
TeV only one side of the contour, the solid line, is visible).}
\label{fig:aztx} 
\end{minipage}
\qquad
\begin{minipage}[t]{0.47\textwidth}
    \vspace{0pt}
    \centerline{\includegraphics[width=\textwidth]{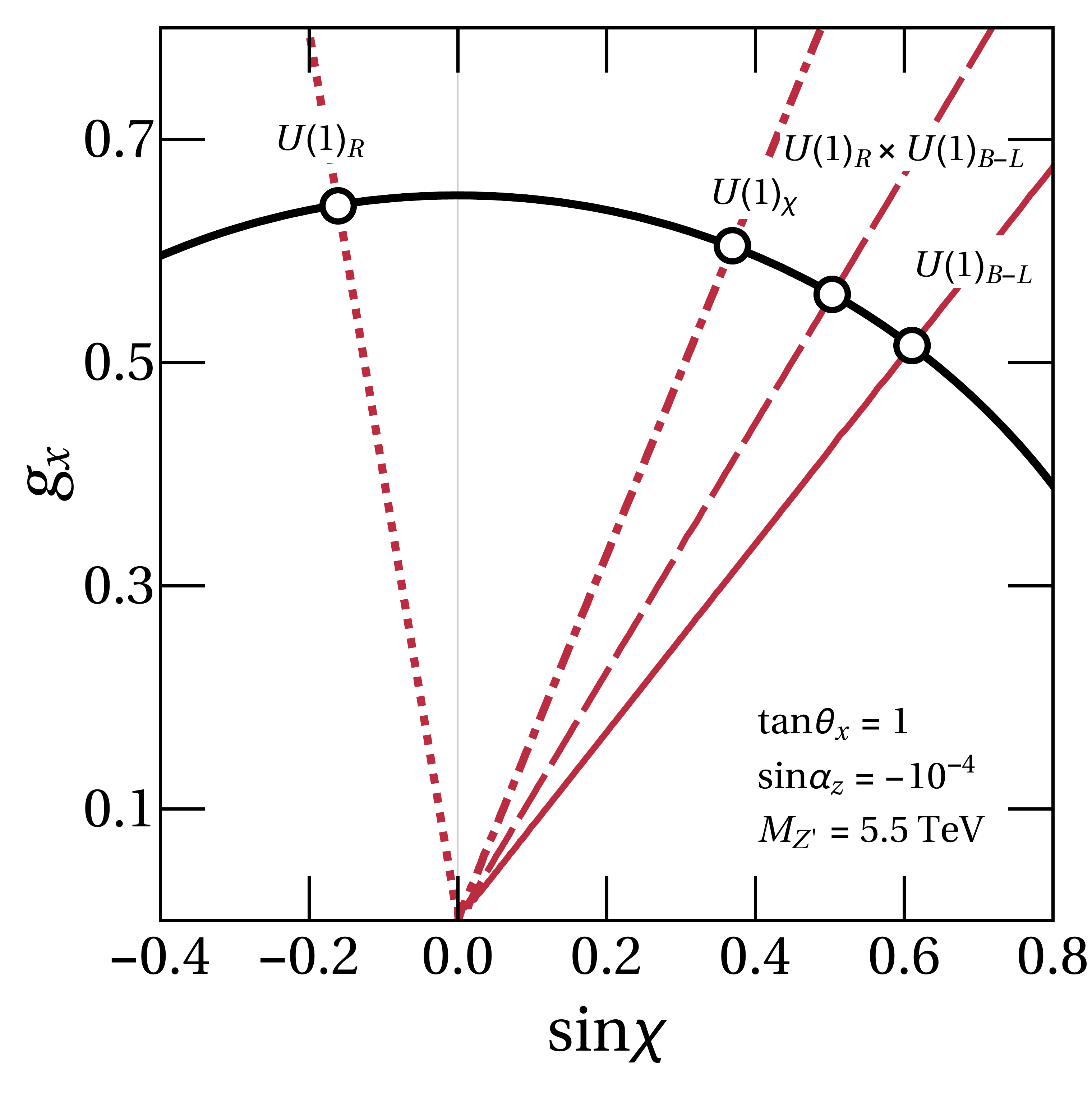}}
    \caption{\small \em Example plot illustrating the inter-relationship
        between kinetic mixing~($\sin\chi$) and the original $U(1)_X$
        coupling~($g_x$) assuming hypothetical measurements:
        $M_{Z'}\approx 5.5$~TeV, $\sin\alpha_z \approx (-10^{-4})$ and
        $\tan\theta_x \approx 1 $. The solid black curve is the contour
        corresponding to \Eqn{e:gzp}.  Each red line corresponds to a
        particular model, drawn in conformity with \Eqs{e:zp}{e:zzp3}.
    The intersection of the black curve with a particular red line gives
the solutions for the kinetic mixing and $g_x$ for a given model.}
\label{fig:hypothetical} 
\end{minipage} 
\end{figure} 
It may be useful to note that every point in the 
$\sin\alpha_z$-$M_{Z'}$ plane in Fig.~\ref{fig:contours}
corresponds, through \Eqn{e:F}, to a definite value of ${\cal F}$.
If a specific model is chosen then one can use the relation
\begin{equation}
\tan\chi = \left(2\kappa_x - \frac{1}{2}\right) \tan\theta_x
\cot\theta_w - \frac{{\cal F}}{\sin\theta_w} \, ,
\label{e:chi}
\end{equation}
which follows from \Eqn{e:zzp3}, to determine the kinetic mixing
angle, $\chi$, corresponding to this point. The value of $\kappa_x$
varies from model to model, $\tan\theta_x$ is a measure of the
effective gauge coupling of the extra $U(1)_X$, and $\cos\theta_w$ is
determined in terms of $\sin\alpha_z$ and $M_{Z'}$ through
\Eqn{e:zzp1}. Conversely, for a fixed value of the kinetic mixing parameter, $\chi$, any
model would correspond to a curve, determined by $\kappa_x$, in the
$\sin\alpha_z$-$M_{Z'}$ plane.  As a definite example, if we
consider the $(B-L)$ model ($\kappa_x = 1/4$), the curve corresponding
to $\chi = 0$ is a vertical straight line through the origin. This is
reminiscent of the fact that in this model $Z$-$Z'$ mixing is entirely
due to kinetic mixing.

In Fig.~\ref{fig:gymz}, we take a complementary approach by casting
the bounds in the $M_{Z'}$-$\tan\theta_x$ plane, for two
representative values of $\sin\alpha_z$, namely, $0$ and
$(-10^{-4})$. For these values of $\sin\alpha_z$ strongest limits come
from direct searches, which have been displayed by the red lines. For
$\sin\alpha_z=0$ the region to the right of the solid red line is
allowed, whereas for $\sin\alpha_z=(-10^{-4})$ the region contained
within the dashed red lines is allowed.  The absence of contours from
considerations of unitarity and $\nu_\mu$-$e$ scattering in
Fig.~\ref{fig:gymz} implies that the corresponding curves are too weak
to enter inside the zoomed range of the parameter space.

In Fig.~\ref{fig:aztx}, we display the bounds in the
$\sin\alpha_z$-$\tan\theta_x$ plane, for two representative values of
$M_{Z'}$, namely, $4$~TeV and $5$~TeV. For these values of $M_{Z'}$
strongest limits come from direct searches, displayed by the red lines.
For $M_{Z'}=4$~TeV the region inside the solid red contour is allowed,
whereas for $M_{Z'}=5$~TeV the region bounded within the dashed red
lines is allowed. The green lines correspond to $\Gamma_{Z'}\le M_{Z'}/2$.

Finally, with the ambitious expectation that a $Z'$ will be
discovered in future, in Fig.~\ref{fig:hypothetical} we
illustrate how model specific information can be extracted using
the following hypothetical measurements of the model-independent
parameters: %
\begin{eqnarray}
M_{Z'} \approx 5.5 \, {\rm TeV} \,, \qquad
\sin\alpha_z \approx (-10^{-4}) \,, \qquad 
\tan\theta_x \approx 1 \,.
\end{eqnarray}
The solid black line in Fig.~\ref{fig:hypothetical} has been obtained
by combining \Eqs{e:tz}{e:gzp} for $\tan\theta_x = 1$. It does not
depend on the chosen model. The red lines, on the other hand, are
drawn using \Eqn{e:zp} in conjunction with \Eqs{e:zzp1}{e:zzp3} to
trade $\theta_w$ and $x'_\Phi$ in favor of $M_{Z'}$, $\alpha_z$ and
$\tan\theta_x$. Since the red lines require the input of $x_\Phi$
which, in turn, depends on $\kappa_x$, the lines are different for
different models. The intersection of the black line with a particular
red line gives the solutions for the kinetic mixing parameter, $\chi$,
and the $U(1)_X$ coupling, $g_x$, for that particular model. Such a
solution might provide intuition as to whether a specific $U(1)_X$
model fits into a more elaborate scheme, such as grand unification, at
higher energies.

%
\section{Conclusions}
Our intention in this paper has been to put constraints on the
parameter space of the minimal extension of the SM with an additional
gauged $U(1)$ giving a massive neutral $Z'$ gauge boson. We did
revisit the formalism first to set up the notations. 
We have advocated a parametrization in which, in the presence of 
kinetic mixing, the constraints on different anomaly-free $U(1)_X$ models
can be expressed in a model-independent unified framework.
Importantly, we have not {\em a priori} assumed, unlike most of the
previous works, that the $Z$-$Z'$ mixing angle is small or the $Z'$ mass
is way above the $Z$ mass.  For the sake of illustration we explicitly
examine a few popular scenarios of $U(1)$ extension, e.g., the $(B-L)$
model, an $U(1)$ arising from left-right symmetry, etc. It turns out
that there are three important quantities to be determined which cover
the extended parameter space and absorb all model dependence for a
non-anomalous $U(1)$ extension. These quantities are the mass of the
$Z'$, the effective gauge coupling strength ($g'_x$) of the extra
$U(1)$, and the $Z$-$Z'$ mixing angle ($\alpha_z$).  To constrain this
space, we have primarily employed three types of information, namely,
the LHC (ATLAS) 13 TeV Drell-Yan data with 36 ${\rm fb}^{-1}$
luminosity, the results from low energy $\nu_\mu - e$ scattering, and
consistency with $s$-wave unitarity in the $W_L^+W_L^- \to W_L^+W_L^-$
channel. The LHC data turn out to be most constraining. {We also
    observe that constraints on the $Z'$ decay width, $\Gamma_{Z'}$,
    translate to constraints in the parameter space which are {\em
    similar in nature} to those obtained from $s$-wave
    unitarity}. {We want to underscore that although we employ the
    anomaly-free (per generation) models to exemplify our formalism, the
    analysis can in general be used to constrain other extensions of the
    SM with an additional $Z'$.  The interplay between the
    different bounds can be used to constrain models with or without
    couplings to fermions, and with or without $Z$-$Z'$ mixing. Also,
    models with a $Z'$ that couples only to leptons, or even
preferentially to the third generation can be constrained using our
study.} The new things that emerge from our analysis are the following:
\begin{itemize}

\item Our parametrization shows that increasingly precise experimental
    data would squeeze the allowed region in the three-dimensional space
    of $M_{Z'}$, $\alpha_z$ and $\theta_x$. The description is
    completely model-independent as long as the fermion content ensures
    an anomaly-free set-up. Model dependence is encoded in $\kappa_x$,
    which is different for different models, as listed in Table
    \ref{t2}.  Of the other parameters, the strength of kinetic mixing,
    $\chi$, should in principle be a derived quantity in a fundamental
    theory given the charges of a possible set of heavy particles
    (couplings both to $B_\mu$ and $X_\mu$), integrated out to generate
    the mixing. Nevertheless, in our approach, which is agnostic towards
    models of UV completion, $\chi$ is treated as an effective
    parameter. Given a model ({\em i.e.} a value of $\kappa_x$), one can
    calculate a range in $\chi$ using \Eqn{e:chi} which would fit values
    (or limits) of $M_{Z'}$, $\alpha_z$ and $\theta_x$ extracted
    directly from experimental data.

\item We have updated the model-independent constraints in the
  $C_u^{\ell}$-$C_d^{\ell}$ ($\ell \equiv e, \mu$) plane, using the
  latest 13 TeV (36 ${\rm fb}^{-1}$) ATLAS data. We obtain an
  improvement of one order of magnitude over the previous constraints
  in the same plane obtained from the publicly available 7-8 TeV CMS
  results \cite{Khachatryan:2014fba} (see also
  \cite{Accomando:2010fz}), and several orders of magnitude over those
  from Tevatron results \cite{Carena:2004xs}. While constraints were
  speculated before actual LHC data arrived
  \cite{Rizzo:2006nw,Petriello:2008zr,Salvioni:2009mt}, our analysis
  provides the most updated ones in the $C_u^{\ell}$-$C_d^{\ell}$
  plane using the latest publicly available LHC (ATLAS)
  data. Translating experimental data to constraints in the above
  plane as a function of $(M_{Z'}, C_u^{\ell}, C_d^{\ell})$, rather
  than directly to limits on $M_{Z'}/g^{'2}_x$, is quite useful as it
  provides a model-independent platform from where limits on any type
  of specific customized models can be easily extracted. ATLAS has
  also provided bounds for Drell-Yan $\tau^+\tau^-$ production through
  a $Z'$. We use this dataset to set similar constraints in the
  $C_u^{\tau}$-$C_d^{\tau}$ plane. Though less restrictive, these
  latter bounds are useful for non-universal $Z'$ models which have a
  different coupling to the third generation fermions.

\item The $s$-wave unitarity constraints in the ($M_{Z'}$-$\sin
  \alpha_z$) plane, placed for the first time in this paper, turn out
  to provide complementary limits when the LHC direct search and the
  low energy $\nu_\mu$-$e$ scattering constraints are superposed in
  the same plane. It is important to observe that the unitarity
  constraints are insensitive to the extra $U(1)$ coupling strength,
  $g'_x$, and in conjunction with the LHC direct search limits they
  restrict the $Z$-$Z'$ mixing to be small (which we have not {\em a
    priori} assumed). 
  However, when we require $\Gamma_{Z'} \le 0.5~M_{Z'}$, the
  constrained turn out to be much stronger than the ones obtained from
  $\nu_\mu$-$e$ scattering data or from satisfying $s$-wave unitarity.
  The constraints on the mixing angle ($\alpha_z$) we obtain are, in
  fact, of the same order as obtained from electroweak precision tests
  \cite{Czakon:1999ha,Erler:2009jh}.

\item When the $Z'$ couples to fermions with the same strength as that
  of the SM $SU(2)_L$ gauge boson (for $(B-L)$ model this corresponds
  to $\tan\theta_x =4$), we obtain $M_{Z'} > 4.4$ TeV and
  $|\alpha_z|<0.001$ at $95\%$ C.L.

\end{itemize}
We urge our experimental colleagues to take notice of our assertion
that a model independent analysis, as depicted especially by the
direct detection contour in Fig.~\ref{fig:contours}, can be carried
out with just three independent parameters, as discussed in detail.


\paragraph*{Note added\,:} While this manuscript was being
finalized, the 13~TeV Drell-Yan data from the CMS Collaboration
became available\cite{Sirunyan:2018exx}. Our result in  the
$C_u^{e,\mu}$-$C_d^{e,\mu}$ plane, which uses the 13~TeV ATLAS
Drell-Yan data, is very similar to that obtained by the CMS
collaboration. Analysis using the 13~TeV ATLAS
Drell-Yan data has also been performed very recently in 
Refs.~\cite{Gulov:2018zij,Pevzner}.


\paragraph*{Acknowledgements\,:} TB acknowledges a Senior
  Research Fellowship from UGC, India.  GB and AR acknowledge support
  of the J.C.\ Bose National Fellowship from the Department of Science
  and Technology, Government of India (SERB Grant
  Nos.\ SB/S2/JCB-062/2016 and SR/S2/JCB-14/2009, respectively).  AR
  also acknowledges suport from the SERB Grant No. EMR/2015/001989.

\renewcommand{\theequation}{A.\arabic{equation}}  
\setcounter{equation}{0}

\section*{Appendix: Detailed expressions for $w_q$}
The NLO expressions for the functions, $w_q$, which appear 
in \Eqn{e:Cq}, are given by
\begin{eqnarray}
    w_q \left(s,M_{Z'}^2\right) &=&  \int_0^1 dx \int_0^1 dy
    \int_0^1 dz \,\, \delta\left(\frac{M_{Z'}^2}{s} - xyz\right) \times
    \nonumber  \\
    &&  \Big\{
    F_{qq}\left(x,y,M_{Z'}^2\right)  \Delta_{qq}\left(z,M_{Z'}^2 \right) + 
    F_{gq}\left(x,y,M_{Z'}^2\right) \Delta_{gq}\left(z,M_{Z'}^2\right)
    \Big\} \,,
 \label{eq:defw}
\end{eqnarray}
For $pp$ colliders such as the LHC we have\cite{Carena:2004xs}:
\begin{subequations}
    \begin{eqnarray}
        F_{qq}\left(x,y,M_{Z'}^2\right) &=& f_{q \leftarrow P}\left(x,M_{Z'}^2\right)
        f_{\bar{q}\leftarrow {P}}\left(y,M_{Z'}^2\right) 
        + \left(x\leftrightarrow y \right) \,,  \\
        F_{gq}\left(x,y,M_{Z'}^2\right)  &=& f_{g\leftarrow
        P}\left(x,M_{Z'}^2\right) \left[
        f_{q \leftarrow {P}}\left(y,M_{Z'}^2\right) + 
                f_{\bar{q}\leftarrow {P}}\left(y,M_{Z'}^2\right) \right]
        + \left(x\leftrightarrow y\right) \,,
    \end{eqnarray}
\end{subequations}
where $f_{q\leftarrow P}\left(x,M_{Z'}^2\right)$ represents the PDF for the
parton $q$ at a factorization scale, $M_{Z'}$.
The scaling functions, $\Delta_{qq}$ and $\Delta_{gq}$, are given by\cite{Hamberg:1990np}
\begin{subequations}
    \begin{eqnarray}
        \Delta_{qq}\left(z,M_{Z'}^2\right) &=& \delta(1-z) + \frac{\alpha_s(M_{Z'}^2)}{\pi} C_F  \left[ \left( \frac{\pi^2}{3}-4 \right) \delta(1-z) -\frac{1+z^2}{1-z}
        \ln(z)  \right. \nonumber \\
        &&  \left. -2 (1+z) \ln(1-z) 
        + 4 (1+z^2)\left(\frac{\ln(1-z)}{1-z}\right)_+\right] \,, \\
        \Delta_{gq}\left(z,M_{Z'}^2\right) &=& \frac{\alpha_s(M_{Z'}^2)}{2\pi}T_F \left[
            (1-2z+2z^2)\ln\frac{(1-z)^2}{z} +\frac{1}{2} + 3z -\frac{7}{2}z^2
        \right] \,,
    \end{eqnarray}
\end{subequations}
where $C_F=4/3$ and $T_F=1/2$ are the quark and gluon color factors
respectively. The plus prescription is defined as follows:
\begin{eqnarray}
    \int_0^1 dx \,\, f(x)\, g(x)_+ = \int_0^1 dx \left[f(x)-f(1) \right] \, g(x) \,.
\end{eqnarray}
We obtained our numerical results using these equations.


\bibliographystyle{JHEP} 
\bibliography{Z-prime}
\end{document}